\documentclass[12pt]{iopart}

\usepackage{graphicx}	
\begin{document}

\title[Slow Relaxation Process in Ising like Heisenberg Kagome Antiferromagnets]
{Slow Relaxation Process in Ising like Heisenberg Kagome Antiferromagnets 
due to Macroscopic Degeneracy in the Ordered State}

\author{Shu Tanaka$^1$ and Seiji Miyashita$^{1,2}$}

\address{
$^1$Department of Physics, University of Tokyo, 7-3-1, Hongo, Bunkyo-ku, Tokyo, 113-0033, Japan

$^2$CREST, JST, 4-1-8 Honcho Kawaguchi, Saitama, 332-0012, Japan
}

\eads{\mailto{shu-t@spin.phys.s.u-tokyo.ac.jp} and \mailto{miya@spin.phys.s.u-tokyo.ac.jp}}

\begin{abstract}
We study relaxation phenomena in the macroscopically ordered state in the Ising like Heisenberg
kagome antiferromagnets. 
In recent experiments, slow relaxation phenomena have been observed in kagome compounds.
We introduce the "{\it weathervane loop}" in order to characterize the
ordered state and study the microscopic mechanism of the slow relaxation 
from a view point of the dynamics of the weathervane loop configuration.
\end{abstract}

\maketitle

\section{Introduction}

Due to the frustration among the magnetic interactions, antiferromagnetic
spin systems on the
triangular, kagome, or pyrochlore, {\it etc}. show very interesting properties of magnetic ordering.
In particular, in the kagome lattice and the pyrochlore lattice, which are 
the so-called corner-sharing lattice where neighboring local triangles  share
a spin but not a bond, frustration causes macroscopic degenerate states in the ground state
even for the continuous spin systems, {\it e.g.} XY, and Heisenberg systems.
Thus, there is no phase transition at finite temperatures in the kagome antiferromagnets
with Ising, XY and Heisenberg spin systems. However, Kuroda and Miyashita pointed out
that the kagome aniferromagnet has a magnetic phase transition with the 
universality class of the ferromagnetic Ising model \cite{Kuroda}, 
when the system has finite easy-axis anisotropy, namely 
the Ising like Heisenberg interaction. There the ordered state
is still macroscopically degenerated, and was called exotic ferromagnetic phase.
Recently slow relaxation phenomena were found in the kagome and pyrochlore lattice 
\cite{Wills1},\cite{Ladieu}.
In this paper we consider the origin of the slow relaxation from the view point of 
structural change of 
the macroscopically degenerate ordered states of the Ising like Heisenberg kagome 
antiferromagnet.

There are various candidate materials for 
the antiferromagnetic kagome spin system, for example,
Rb$_2$M$_3$S$_4$ \cite{Wills1}-\cite{Kuki} (M
is a magnetic ion such as Ni, Co, Mn), AM$_{3}$(OH)$_{6}$(SO$_{4}$)$_{2}$, (A$^{+}$ is a cation such as
K$^{+}$, Rb$^{+}$, NH$_{4} ^+$, Tl$^+$, Ag$^+$ or Na$^+$, and M$^{3+}$ is a magnetic
ion such as Fe$^{3+}$, or Cr$^{3+}$) \cite{Maegawa}-\cite{Inami}, 
NaFe$_{3}$(SeO$_{4}$)$_{2}$(OH)$_6$ \cite{Oba}, SrCr$_{9x}$Ga$_{12-9x}$O$_{19}$
(SCGO) \cite{Ladieu}, 
and ${}^3$He on the sheet of graphite \cite{Graywall}.
Maegawa {\it et.al.} discovered the two cusps of the temperature dependence of the
susceptibility in NH$_4$Fe$_3$(OH)$_6$(SO$_4$)$_2$ \cite{Maegawa}. They
concluded that the successive phase transition may be caused by a small
Ising-like anisotropy in the Heisenberg kagome antiferromagnet. This result is
supporting Kuroda and Miyashita's scenario. 
In recent experimental studies, the antiferromagnetic kagome compounds shows
the slow relaxation of magnetization and magnetic
susceptibility. Usually, the slow dynamics is caused by random interaction
of the systems. However, the slow relaxation appears in non-random
spin systems such as SrCr$_{9x}$Ga$_{12-9x}$O$_{19}$ (SCGO), 
AM$_{3}$(OH)$_{6}$(SO$_{4}$)$_{2}$ which are corner sharing structure.

In the present paper, we study relaxation phenomena in kagome antiferro magnetic systems 
in the macroscopically degenerate ordered state. 
In order to characterize the degenerate state we introduce
"weathervane loop", and investigate the microscopic mechanism of 
the slow relaxation in kagome antiferromagnetic systems by considering relaxation
of configuration of the weathervane loop structure, which is much slower than the 
relaxation of the total magnetization which is the order parameter of the model. 

In Section 2, we review the phase transition and
features of the Ising like Heisenberg antiferromagnetic kagome systems. 
In Section 3, we consider relaxation processes of 
the number of 'weathervane loop'.
In Section 4, we conclude our research. 

\section{Model}

We consider the antiferromagnetic Ising like Heisenberg kagome antiferromagnetic system, 
\begin{equation}
 \mathcal{H} = J \left(\sum _{\left\langle i,j \right\rangle} S_i ^x S_j ^x +
  S_i ^y S_j ^ y + A S_i ^z S_j ^z\right), \quad J>0  \,\,{\rm and} \,\,\, A>1,
\end{equation}
where $\left\langle i,j \right\rangle$ denotes nearest neighbor in
kagome lattice and $A$ denotes Ising like anisotropy. Kuroda and
Miyashita have shown that the system has a phase transition which
belongs to the universality class of two dimensional ferromagnetic Ising
spin system \cite{Kuroda}. The kagome system consists of triangle unit
that shares one corner. The ground state of Ising like Heisenberg
triangle unit is given by
$S_\alpha = \left( 0,0,1 \right)$, 
$S_\beta =\left( s,0,-c \right)$, $S_\gamma = \left( -s,0,-c \right)$, 
where $c=\frac{A}{A+1}$ and $s=\sqrt{1-c^2}$. 
The freedom of the rotation $2\pi$ in the $xy$ plane remains in $S_\beta$ and $S_\gamma$.

It should be noted that this system has nonzero magnetization in the ground state,
the value of which
is $1-2c$ in each triangle unit. 
The magnetization of the ground state per spin is 
$\frac{M_0}{N} = m_0 = \frac{1}{3} \times \left( 1-2c \right)$,
where $M_0$ and $N$ denote the total magnetization of the ground state 
and the number of spins, respectively.
It is important to note that no sublattice long range order 
in this system in spite of the existence of the magnetic phase transition.

In order to study the equilibrium properties, 
we use the heat bath method of Monte Carlo simulations 
with $100,000$ Monte Carlo Steps(MCS) for initial relaxation and 
$100,000$ MCS for collecting data.
For the region near the critical point, we perform $200,000$ MCS for initial relaxation and 
$500,000$ MCS for measurement.
Figures 1 show temperature dependence of (a) the magnetization 
and (b) the specific heat for $A=3$.

\begin{figure}[h] 
 \begin{center}
 \includegraphics[height=6cm,clip]{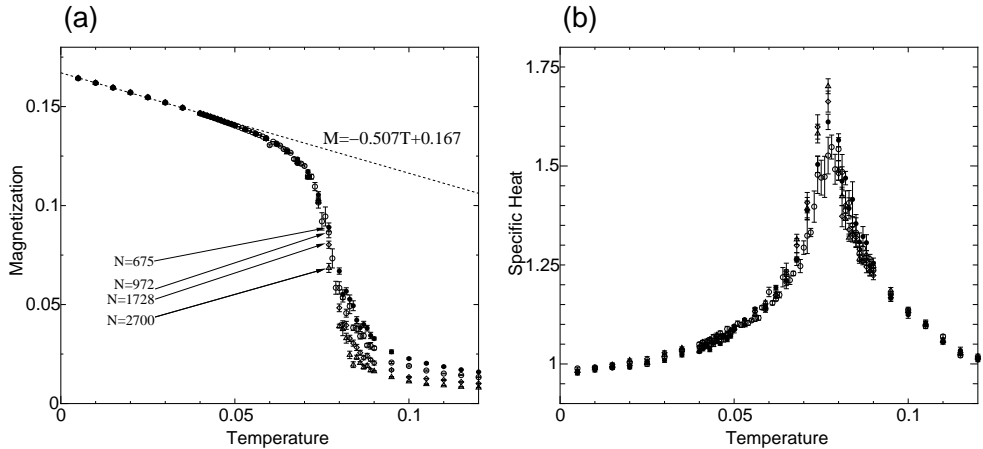}
\caption{Temperature dependence of (a) the magnetization and 
(b) the specific heat for $A=3$. (\fullcircle  for $N=675$, \opencircle  for $N=972$,
\opendiamond  for $N=1728$, and \opentriangle  for $N=2700$)}
 \end{center}
\end{figure}

When $T \to 0$, the magnetization approaches the ground state $m_0 = -\frac{1}{6}$.
At $T \simeq 0.078$, the magnetization changes suddenly and the specific heat diverges,
which indicates the second order phase transition. 
This is the phase transition which belongs to the two dimensional 
Ising ferromagnetic universality class \cite{Kuroda}. 

\section{Weathervane Loop and Defects}

Because the antiferromagnetic kagome lattice system has large number of
degenerated states, it is important to consider the
entropy of the spin configuration. If there is the easy-axis type anisotropy ({\it i.e.} $A>1$), 
one spin ($S_\alpha$) is parallel to the $z$-axis and two spins ($S_\beta$,$S_\gamma$) 
direct opposite direction in each triangle cluster in the ground state.
The spins $\{ S_\beta \}$ and $\{ S_\gamma \}$ have 
the freedom of the $2\pi$ rotating in the $xy$ plane.
If we connect $\{ S_\beta \}$ and $\{ S_\gamma \}$ in the lattice, 
we find a closed loop as shown in Fig. 2.
We call the line {\it weathervane loop}. 

\begin{figure}[h]
 \begin{center}
  \includegraphics[height=2.5cm,clip]{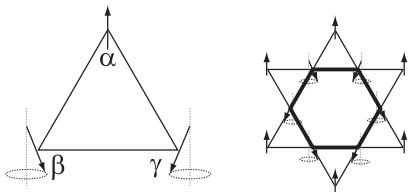}
  \caption{The spin configuration of the ground state in a triangle cluster and an example
  of the weathervane loop (bold line).}
 \end{center}
\end{figure}

In the ground state, there are macroscopic degenerate configurations of the weathervane loop 
in Fig. 3(a),(b), and (c).
Figs. 3(a) and (b) show the ground state which are called 
$q=0$ and $\sqrt{3}\times\sqrt{3}$ structure, respectively.
On the other hand, in Fig. 3(c), we show one which obtained by 
quenching the system from high temperature.
This is one of the degenerate configuration and we call it "random structure".
Let us consider the degeneracy of the ground state.
Each loop has the rotation degree of freedom $2\pi$. 
We denote the number of the degeneracy of the ground state $\left( 2\pi \right)^{n_{\mathrm{loop}}}$, 
where $n_\mathrm{loop}$ denotes the number of weathervane loop.
Here we take the degeneracy of single angle to be $2\pi$.
In the spin configuration of the $\sqrt{3} \times \sqrt{3}$ structure, 
$n_{\mathrm{loop}}$ takes the maximum value $n_{\mathrm{max}}=N/9$. 

\begin{figure}[h]
 \begin{center}
  \includegraphics[height=4cm,clip]{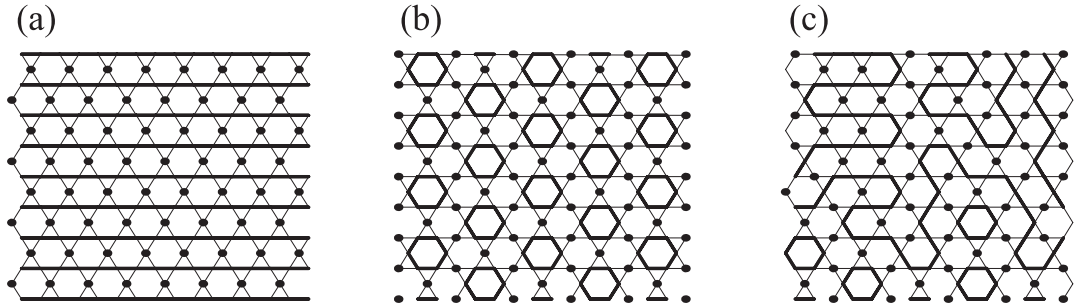}
  \caption{The typical example of the ground state of the anisotropic Heisenberg kagome system. 
  (a)$q=0$ state, (b)$\sqrt{3}\times\sqrt{3}$ state, and (c)the random ground state. The circles and 
  thick lines denote the $S_\alpha$ and the weathervane loop, respectively.}
 \end{center}
\end{figure}

Next, we consider the time evolution of the number of the weathervane loop and the magnetization.
Here we study the system of $A=3$ and $N=2700$.
In Fig. 4(a), we show time evolution of the magnetization at $T=0.07$ 
starting from the initial three types of spin states, {\it i.e.} random, $q=0$, 
and $\sqrt{3}\times\sqrt{3}$.
Here the data are obtained as an average over 36 independent runs. 
The errorvars are of order 0.05 which are omitted.
Though the magnetization relaxes to the equibrium value in a time, 
$\tau_{\mathrm{mag}}^{\left( T=0.07 \right)} \sim 10^3$ MCS,
it takes $\tau_{\mathrm{loop}} \sim 10^5$ MCS for $n_{\mathrm{loop}}$ to relax (Fig. 4(c)).
In the inset the initial part is magnified, 
we find the different relaxation processes 
in the three cases. In the case of random pattern, we find a simple exponential type relaxation.
From $\sqrt{3} \times \sqrt{3}$ structure, the $n_\mathrm{loop}$ reduces from the maximum value to 
the equilibrium value monotonically. On the other hand, from $q=0$ structure, $n_\mathrm{loop}$ shows an 
non-monotonic behavior.

At $T = 0.05$, the relaxation time of the magnetization is 
$\tau_{\mathrm{mag}}^{\left( T=0.05 \right)} \sim 10^3$ MCS same as the case of $T=0.07$(Fig. 4(b)).
In this case, it should be noted that the relaxation time of 
the weathervane loop is more than $10^7$ MCS(Fig. 4(d)). 
This slow relaxation is attributed to the cost that is necessary for the re-arrangement of
the weathervane loop. This cost is large comparing to the temperature.
The difficulty of the re-arrangement of the weathervane loop is the origin of the slow relaxation 
in easy-axis type anisotropic Heisenberg kagome antiferromagnets.

\begin{figure}[h]
 \begin{center}
  \includegraphics[height=11cm,clip]{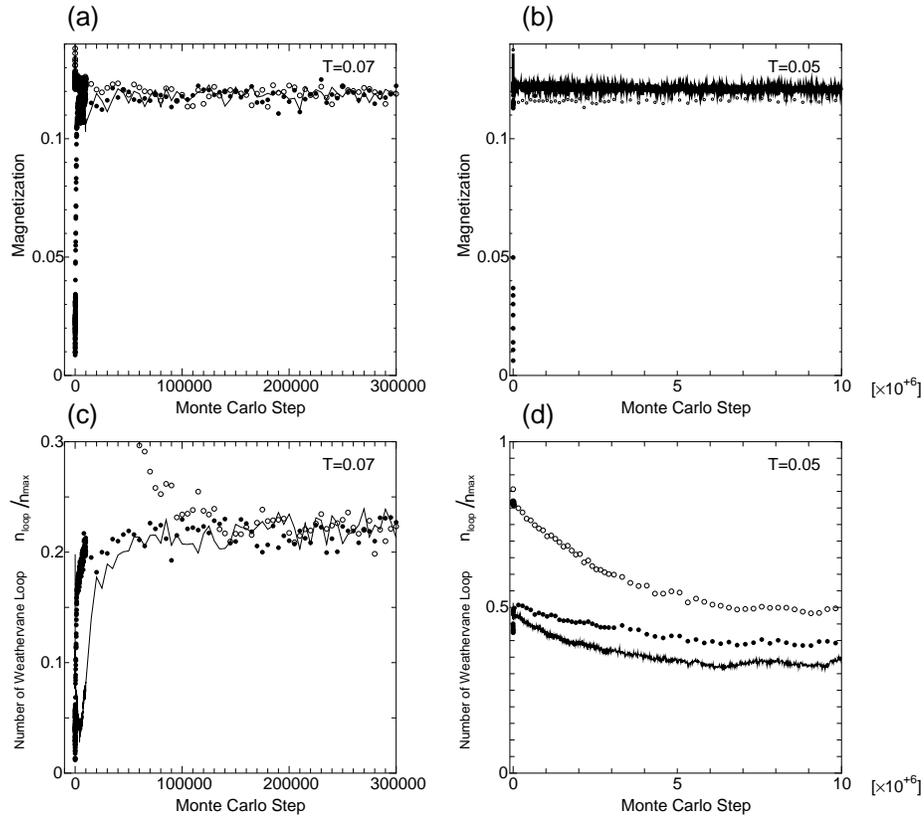}
  \caption{Time evolution of the magnetization (a) at $T=0.07$ and (b) at $T=0.05$, 
  and $n_{\mathrm{loop}}$ (c) at $T=0.07$ and (d) at $T=0.05$.
          Symbols denote the initial conditions as (\opencircle for $\sqrt{3}\times\sqrt{3}$,
  a solid line for $q=0$, and \fullcircle for the random).}
 \end{center}
\end{figure}

\section{Conclusion}

We consider the origin of the slow dynamics in anisotropic Heisenberg kagome antiferromagnets. 
In this study, the main stress falls on the difficulty of the re-arrangement of the weathervane loop, 
which is the origin of the slow relaxation in this system.
In this paper, we consider two dimensional kagome antiferromagnetic classical spin systems 
with Ising like anisotropic interaction. 
Below the critical temperature, the relaxation of the $n_{\mathrm{loop}}$ is slower than of the
magnetization which is the order parameter of this model.
We expect to realize our scenario in the easy-axis type anisotropic Heisenberg antiferromagnets 
in the ferromagnetic ordered state. When the inter-plane interaction exists, we have to 
study the present mechanism in three dimensions.
There the weathervane loop becomes weathervane plane and the recombination will
have more significant effect on the relaxation, which will be reported elsewhere \cite{Tanakafuture1}.

\ack
The present work is partially supported by Grand-in-Aid from the Ministry of Education, 
Culture, Sports, Science, and Technology, 
and also by NAREGI Nanoscience Project, Ministry of Education Culture, Sports, Science, and Technology, Japan.

\end{document}